\documentclass{article}
\newcommand{\cc}{\cite}

\newcommand{\be}{\begin{equation}}
\newcommand{\ee}{\end{equation}}

\def\ve{\varepsilon}
\def\w{\omega}

\def\pd{\partial}

\def\L{\Lambda}

\def\z{\zeta}

\def\<{\langle}
\def\>{\rangle}

\def\Log{\hbox{ln}}

\def\g{\gamma}  \def\G{\Gamma}
  
   \def\L{\Lambda}

\def\r{\rho}

\def\m{\mu}
\def\n{\nu}

\def\z{\zeta}
\def\w{\omega}

\def\v{\vec}

\def\({\left(}
\def\[{\left[}
\def\){\right)}
\def\]{\right]}

\def\pd{\partial}
\def\dk{{d^3 k \over (2\pi)^3}}

\def\w1{W^{(1)}}
\def\v1{V^{(1)}}

\begin{document}

\begin{center}
{\Large \bf On Casimir Energy Contribution to Observable Value
\\ of the Cosmological Constant} \\

\vspace{10mm}

Igor O. Cherednikov\footnote{Email: igorch@thsun1.jinr.ru} \\
\vspace{5mm}
\it Bogolyubov Laboratory of Theoretical Physics, \\
Joint Institute for Nuclear Research \\
141980 Dubna, Russia\\
and \\
Institute for Theoretical Problems of Microphysics, \\
Moscow State University \\ 119899 Moscow, Russia
\\
and \\
The Abdus Salam International Centre for Theoretical Physics, \\
34100 Trieste, Italy
\\
\end{center}

\begin{abstract}
\noindent
The contribution of the ground state energy of quantum fields to the cosmological
constant is estimated from the point of view of the standard Casimir energy calculation
scheme. It is shown that the requirement of the
renormalization group invariance leads to the value of the effective $\L$-term
which is of 11 orders higher than the result extracted from the experimental
data.  \end{abstract}
\vspace{5mm}

\section{Introduction}
\noindent
One of the most intriguing and challenging problems of the modern physics is
the enormously large difference between the experimentally extracted
cosmological constant and it's value estimated within the convenient quantum
field theory \cc{rev}. This is so-called old cosmological constant problem,
while the new one is why it is comparable to the present mass density
\cc{wei, vil}.  Here we will study only the old one.  This problem attracts a
great interest of the physical community (just mention that the SPIRES-SLAC
database gives more than 800 references for ``cosmological constant''), and a
lot of sophisticated approaches have been proposed to it's solution \cc{rev, star}.
However, the present situation seems to be far from satisfactory.

The Einstein's equation
\be
R_{\m\n} - {1\over2} g_{\m\n} R = g_{\m\n} \L- 8 \pi G T_{\m\n} \ ,
\label{ei} \ee where $G$ is the gravitational constant and other notation are
standard \cc{rev}, contains the classical $\L$-term as well as the
contribution of the vacuum energy density due to the quantum fluctuations, which had been
shown by Zeldovich \cc{zel1} to have the same structure:
$\<T_{\m\n}\> = - g_{\m\n} \<\r\>$. It means that  (\ref{ei}) can be written
in the form:  \be
R_{\m\n} - {1\over2} g_{\m\n} R = g_{\m\n} \L_{eff} \ ,
\ee where
\be
\L_{eff} = 8 \pi G \r_{eff} = \L + 8 \pi G \<\r\> \ . \label{den}
\ee It is known from the experimental data that the effective energy density of
the Universe $\r_{eff}$ defined in (\ref{den}), is of order $10^{-47} GeV^4$ \cc{ast}.
In the same time, the directly evaluated vacuum energy with the UV cutoff at
the Planck scale $M = \(8\pi\)^{-1/2} m_P$, $m_P \approx 1.2210(9)\ 10^{19} GeV$ \cc{pdg}
reads
\be
\<E_{vac}\> = {1 \over 2} \ \int_0^M \! \dk \sqrt{k^2 + m^2} \approx {M^4 \over 16 \pi^2} \ ,
\ee is about $10^{71} GeV^4$, {\it i. e.,} 118 orders higher. Even
if one takes the cutoff at the supersymmetry breaking scale
$\r_{eff}^{susy} \sim 10^{12} GeV^4$, the discrepancy will remain to be
of 59 orders \cc{rev, vil}.

In the present paper, we propose to calculate the vacuum energy $\<\r\>$
within the framework of the standard Casimir energy computations for various
geometrical configurations with quantized fields under boundary conditions.
This approach allows one to obtain the finite value for the Casimir energy by
means of absorption the singularities into the definitions of the
corresponding classical contact terms which characterize the total energy of
the analogous classical configuration \cc{cas}. In our case, the divergences
will be absorbed into the definition of the single ``classical'' parameter $\L
\to \L_0$ which is treated therefore as a ``bare'' constant from the
beginning.

\section{Renormalization of the Casimir energy contribution}
\noindent
Let us consider the ``toy'' Universe filled with the free neutrino field
with the mass $m_\n$ of order $10^{-9} GeV$. The crusial role of light neutrinos in
generating the small non-vanishing value of the cosmological term
had been proposed and discussed in \cc{shap} in the case of spontaneous symmetry
breaking (SSB). The total energy of the vacuum fluctuations of this field reads
\be
E_\n = - \int \! \dk \sqrt{k^2 + m_\n^2} \ . \ee
This integral diverges and then must be regularized before any calculation
will be done. We use the $\z$-regularization which seems to be one of the most
convenient for this situation. Besides this, we need to introduce the
additional mass parameter $\m$ (such an arbitrary mass scale emerges
unavoidably in any regularization scheme) in order to restore the correct
dimension for the regularized quantities.
Then we have
\be \<\r\> = E_\n \to E_\n (\ve)= - \m^{2\ve} \ \int \! \dk \ {1 \over
\(k^2 + m_\n^2\)^{\ve - {1 \over 2}} } \ . \ee The regularization will be removed by
taking the limit $\ve \to 0$. After the simple calculations, we get
\be E_\n  (\ve) = - {m_\n^4 \over 8 \pi^{3/2}} \ \({\m^2
\over m_\n^2}\)^{\ve} \ {\G(\ve -2) \over \G(\ve - {1 \over 2})} \ . \ee
Taking into account the well-known relations for the $\G$-function \cc{int}
\be
\G(\ve -2) = {\G(1+\ve) \over \ve(\ve-1)(\ve-2)} \ \ , \ \ \G(\ve -1/2)
= {\G\({1 \over 2} + \ve\) \over \ve - {1 \over 2}} \ \ , \ee and the
expansions for small $\ve$ $$ \G(1+\ve) = 1 -\g_E\ve + O(\ve^2)\ \ , \ \ x^\ve
= 1 + \ve \ \Log x + O(\ve^2)\ \ , $$ \be   \G\({1 \over 2} + \ve\) = \G\({1
\over 2}\) - \ve \G\({1 \over 2}\)(\g_E + 2 \Log 2) + O(\ve^2) \ ,  \ee we
find:  \be E_\n (\ve) = {m_\n^4 \over32 \pi^2} \({1 \over \ve } + 2 \Log 2 -
{1 \over 2}\)  \ \Log \({\m^2 \over m_\n^2}\) + O(\ve) \ . \ee The part
singular in the limit $\ve \to 0$ can be extracted as  \be E_{div} (\ve) = {1
\over \ve }\ {m_\n^4 \over32 \pi^2} \ .  \ee The absence of the leading divergence
scaled as $m_P^4$ is the generic feature of the $\z$-function regularization
method, and is well-known in the Casimir energy calculations \cc{cas, cas1, rg}.
Thus, the renormalization is performed via the absorption of this singularity
into the re-definition of the bare classical constant $\L$:
\be
\L \to \L_0 - {1 \over \ve }\ {m_\n^4 \over32 \pi^2} \ .
\ee
Therefore, the remaining finite value for the effective energy density reads
\be
\r_{eff} = {\L_0 \over 8 \pi G} + {m_\n^4 \over32 \pi^2} \( \Log
\({\m^2 \over m_\n^2}\) + 2\Log 2 - {1 \over 2} \) \ . \label{r1}
\ee
This quantity depends on the arbitrary mass scale $\m$. It is natural to demand
it to be unchanged under any variations of this parameter. The role of such a
condition in the Casimir energy calculations have been studied in  \cc{cas, rg},
and investigated in detail in \cc{shap, new}.
This requirement leads to the renormalization group equation \be
\m {d \over d\m} \r_{eff} = {1 \over 8 \pi G} \ \m {\pd \over \pd \m} \L_0
(\m) + {m_\n^4 \over 16 \pi^2} = 0 \ . \ee
Solving it we find that the renormalized constant $\L_0$ should be treated as
a ``running'' one in that sense that it varies provided that the scale $\m$ is
changing:
\be
\L_0 (\m) = - {G m_\n^4 \over 2 \pi} \Log {\m \over \m_0} \ , \label{rgs1}
\ee where $\m_0$ can be called the normalization point, determined by the
condition  \be \L_0 (\m_0) = 0  \ . \label{norma}\ee
Substituting (\ref{rgs1}) into (\ref{r1}) we find
\be
\r_{eff} = {m_\n^4 \over16 \pi^2} \( \Log
{\m_0 \over m_\n} + 2\Log 2 - {1 \over 2} \) \ .
\ee
If we assume that the normalization point $\m_0$ coincides with the Planck
scale, {\it i. e.,}  $\m_0 = m_P \sim 10^{19}GeV$, and take the light neutrino mass
to be $m_\n \sim 10^{-9} GeV$, the estimate for the total effective
renormalized cosmological constant will read:
\be
\r_{eff} \sim 10^{-36} GeV^4 \ .  \label{value}\ee

\section{Conclusion}
\noindent
We see that by virtue of the normalization condition (\ref{norma}), one obtains now
the model of universe with $\L_0 = 0$ at the energy scale compared to that in the first moments
of it's existence. The other possible normalization --- $\L_0 = 0$ in the very far IR region
is used and discussed in \cc{shap}.

This value (\ref{value}) is still far from the experimentally observed one, but is much
better that a straightforward evaluation of the ground state energy of quantum
field based on the direct UV cutoff. It should be mentioned that this result is close in order
to one obtained by Ya. B. Zeldovich $(\r_{Zel} \sim 10^{-38}GeV^4)$ by means of the
quite different considerations \cc{zel2}.

One can see that the value of the cosmological constant depend crucially from
the chosen mass of the elementary fermion $m_\n$ and, in contrast, the
dependence from the normalization point $\m_0$ is only logarithmic and may be
neglected, in contrast to the direct evaluation based on the UV cutoff of the
high frequency contributions of the quantum field fluctuations.

\vspace{5mm}

\section{Acknowledgements}
\noindent
The author thanks Prof. I. L. Shapiro for drawing attention to his recent papers on
the similar topics, and Dr. H. Stefancic for the discussion.
This work is partially supported by RFBR (grants Nos. 01-02-16431 and
00-15-96577), and INTAS (grant No. 00-00-366). The warm hospitality of
the Abdus Salam ICTP in Trieste during Oct - Dec 2001 is also thanked.

\end{document}